\documentclass[aps,prd,showpacs,twocolumn,superscriptaddress]{revtex4-1}
\usepackage{graphicx}
\usepackage{epsfig,graphics,subfigure,psfrag,amsmath,amssymb}
\usepackage{lineno}
\usepackage{dcolumn}
\usepackage{bm}
\usepackage{overpic}
\usepackage{xspace}
\usepackage{rotating}
\usepackage{color}
\usepackage{multirow}
\usepackage{colortbl}
\usepackage{epstopdf}
\usepackage[colorlinks,linkcolor=blue,anchorcolor=blue,citecolor=blue]{hyperref}
\usepackage[table]{xcolor}

\newcommand{\ra}{\rightarrow}

\newcommand{\jpsi}{J/\psi}
\newcommand{\jp}{J/\psi}

\newcommand{\pip}{\pi^{+}}
\newcommand{\pim}{\pi^{-}}
\newcommand{\pippim}{\pi^{+}\pi^{-}}
\newcommand{\etap}{\eta^{\prime}}

\newcommand{\gammapipi}{\gamma\pi^{+}\pi^{-}}
\newcommand{\pipieta}{\pi^{+}\pi^{-}\eta}
\newcommand{\etac}{\eta_{c}}
\newcommand{\gam}{\gamma}

\newcommand{\ppe}{\pi^{+}\pi^{-}\eta}
\newcommand{\pp}{\pi^{+}\pi^{-}}

\newcommand{\doubg}{\gamma\gamma}
\newcommand{\doube}{\eta\eta}
\newcommand{\BESIII}{BES\uppercase\expandafter{\romannumeral3}\xspace}

\begin{document}
\title{\boldmath Search for the \texorpdfstring{$X(2370)$}{Lg} and observation of \texorpdfstring{$\etac\to\eta\eta\eta^\prime$}{Lg} in \texorpdfstring{$J/\psi\to\gamma  \eta\eta\etap$}{Lg}}

\author{
M.~Ablikim$^{1}$, M.~N.~Achasov$^{10,c}$, P.~Adlarson$^{64}$, S. ~Ahmed$^{15}$, M.~Albrecht$^{4}$, R.~Aliberti$^{28}$, A.~Amoroso$^{63A,63C}$, Q.~An$^{60,47}$, X.~H.~Bai$^{54}$, Y.~Bai$^{46}$
% ...
\\
\vspace{0.2cm}
(BESIII Collaboration)\\
\vspace{0.2cm} {\it
$^{1}$ Institute of High Energy Physics, Beijing 100049, People's Republic of China\\
% ...
$^{4}$ Bochum Ruhr-University, D-44780 Bochum, Germany\\
% ...
$^{10}$ G.I. Budker Institute of Nuclear Physics SB RAS (BINP), Novosibirsk 630090, Russia\\
% ...
$^{15}$ Helmholtz Institute Mainz, Johann-Joachim-Becher-Weg 45, D-55099 Mainz, Germany\\
% ...
$^{28}$ Johannes Gutenberg University of Mainz, Johann-Joachim-Becher-Weg 45, D-55099 Mainz, Germany\\
% ...
$^{46}$ Southeast University, Nanjing 211100, People's Republic of China\\
$^{47}$ State Key Laboratory of Particle Detection and Electronics, Beijing 100049, Hefei 230026, People's Republic of China\\
% ...
$^{54}$ University of Jinan, Jinan 250022, People's Republic of China\\
% ...
$^{60}$ University of Science and Technology of China, Hefei 230026, People's Republic of China\\
% ...
$^{63}$ University of Turin and INFN, (A)University of Turin, I-10125, Turin, Italy; (B)University of Eastern Piedmont, I-15121, Alessandria, Italy; (C)INFN, I-10125, Turin, Italy\\
$^{64}$ Uppsala University, Box 516, SE-75120 Uppsala, Sweden\\
% ...
$^{c}$ Also at the Novosibirsk State University, Novosibirsk, 630090, Russia\\
% ...
}
}
\date{\today}

\begin{abstract}
Using a sample of $1.31\times10^{9} ~J/\psi$ events collected with the BESIII detector, we perform a study of $J/\psi\to\gamma\eta\eta\etap$ to search for the $X(2370)$ and $\etac$ in the $\eta\eta\etap$ invariant mass distribution. 
No significant signal for the $X(2370)$ is observed, and we set an upper limit for the product branching fraction of ${\cal B}(\jp\to\gamma X(2370)\cdot{\cal B}(X(2370)\to\eta\eta\etap) < 9.2\times10^{-6}$ at the 90\% confidence level.
A clear $\etac$ signal is observed for the first time, yielding a  product branching fraction of ${\cal B}(\jp\to \gamma \etac)\cdot{\cal B}(\etac\to \eta\eta\etap) = (4.86\pm0.62~({\rm stat.})\pm0.45~({\rm sys.}))\times10^{-5}$.
\end{abstract}

\maketitle

\section{INTRODUCTION}
The non-Abelian property of quantum chromodynamics (QCD) permits the existence of glueballs formed by gluons, the gauge bosons of the strong force~\cite{bibg1,bibg2,bibg3}. 
The search for glueballs is an important field of research in hadron physics.
However, the identification of glueballs is difficult in both experiment and theory due to the possible mixing of the pure glueball states with nearby $q\bar{q}$ nonet mesons.
Lattice QCD (in the quenched approximation) predicts the lowest-lying glueballs are scalar (mass 1.5$-$1.7 GeV/$c^2$), tensor (mass 2.3$-$2.4 GeV/$c^2$), and pseudoscalar (mass 2.3$-$2.6 GeV/$c^2$)~\cite{UKQCD,MNstar,bib3,Gregory:2012hu,Sun:2017ipk}.

The radiative decay $J/\psi \to \gamma g g$ is a gluon-rich process and is therefore regarded as one of the most promising hunting grounds for glueballs~\cite{bibjpsi1,bibjpsi2}.
A possible pseudoscalar glueball candidate, the $X(2370)$, is observed in the $\pi^{+}\pi^{-}\etap$ invariant mass distribution through the decays of $J/\psi\to\gamma\pi^{+}\pi^{-}\etap$~\cite{PRL1} and in the $K\bar{K}\etap$ invariant mass distribution in the decays of $J/\psi\to\gamma K\bar{K}\etap$~\cite{kketap} with statistical significances of 6.4$\sigma$ and 
8.3$\sigma$, respectively. 
The measured mass is consistent with the LQCD prediction for the pseudoscalar glueball~\cite{bib3}. 
In a calculation using an effective Lagrangian that couples the pseudoscalar glueball to scalar and pseudoscalar mesons, the ratios of the branching fractions of the pseudoscalar glueball decays $\Gamma_{G\to \eta\eta\etap}$/$\Gamma^{\rm tot}_{G}$, $\Gamma_{G\to KK\etap}$/$\Gamma^{\rm tot}_{G}$ and $\Gamma_{G\to \pi\pi\etap}$/$\Gamma^{\rm tot}_{G}$ are predicted to be 0.00082, 0.011 and 0.090~\cite{PRD1}, respectively, for an assumed glueball mass of 2.370~GeV/$c^2$.  
An observation of the $X(2370)$ in $\jpsi\ra\gam\eta\eta\etap$ would contribute 
to our understanding of this state.  
In parallel, we search for the $\etac$ since this charmonium state has never been observed decaying to $\eta\eta\etap$~\cite{pdg}.  

In this paper, the $X(2370)$ and $\etac$ are studied via $J/\psi\to\gamma \eta\eta\etap$ using $(1310.6 \pm  7.0) \times 10^6$ $J/\psi$ decays~\cite{jpsinumber} collected with the BESIII detector in 2009 and 2012.
The $\etap$ is reconstructed via the decay channels $\etap \ra \gam \pp$ and $\etap \ra \ppe$, and $\eta$ via the decay channel $\doubg{}$.

\section{DETECTOR AND MONTE CARLO SIMULATIONS}
The BESIII detector is a magnetic spectrometer~\cite{Ablikim:2009aa} located at the Beijing Electron Positron Collider (BEPCII)~\cite{Yu:IPAC2016-TUYA01}. 
The cylindrical core of the BESIII detector consists of a helium-based  multilayer drift chamber (MDC), a plastic scintillator time-of-flight system (TOF), and a CsI(Tl) electromagnetic calorimeter (EMC), which are all enclosed in a superconducting solenoidal magnet providing a 1.0~T (0.9~T in 2012) magnetic field. 
The solenoid is supported by an octagonal flux-return yoke with resistive plate counter muon identifier modules interleaved with steel. 
The acceptance of charged particles and photons is 93\% over the $4\pi$ solid angle.  
The charged-particle momentum resolution at $1~{\rm GeV}/c$ is $0.5\%$, and the $dE/dx$ resolution is $6\%$ for the electrons from Bhabha scattering. 
The EMC measures photon energies with a resolution of $2.5\%$ ($5\%$) at $1$~GeV in the barrel (end cap) region.
The time resolution of the TOF barrel part is 68~ps, while that of the end cap part is 110~ps.

Simulated samples produced with the {\sc geant4}-based~\cite{geant4} Monte Carlo (MC) package which includes the geometric description of the BESIII detector and the detector response, are used to determine the detection efficiency and to estimate the backgrounds.
The simulation includes the beam energy spread and initial state radiation (ISR) in the $e^+e^-$ annihilations modeled with the generator {\sc kkmc}~\cite{ref:kkmc}.
The inclusive MC sample includes production of the $J/\psi$ resonance as well as continuum processes incorporated with {\sc kkmc}~\cite{ref:kkmc}.
The known decay modes are modeled with {\sc evtgen}~\cite{ref:evtgen} using branching fractions taken from the Particle Data Group (PDG)~\cite{pdg}, and the remaining unknown decays from the charmonium states with {\sc lundcharm}~\cite{ref:lundcharm}. 
Final state radiation (FSR) from charged final state particles is incorporated with the {\sc photos} package~\cite{photos}.
To estimate the selection efficiency and to optimize the selection criteria, signal MC events are generated for $\jp\to\gamma X(2370),\gamma\etac\to\gamma\eta\eta\etap$. The polar angle of the photon in the $\jp$ center of mass system, $\theta_{\gamma}$, follows a $1+ \mathrm{cos}^{2}\theta_{\gamma}$ distribution. 
The decay of $X(2370)/\etac\to\eta\eta\etap$ is simulated using phase-space (PHSP) generator. So does the process $\etap\to\eta\pp$.
To obtain the efficiency curves, MC events are generated for $\jpsi\to\gam X,\ X\to\eta\eta\etap$, where $X$ means $0^{-+}$ non-resonant state.
For the process $\etap\to\gamma\pi^{+}\pi^{-}$, a generator taking into account both the $\rho - \omega$ interference and the box anomaly is used~\cite{gammapipiDIY}.
The analysis is performed in the framework of the BESIII offline software system (BOSS)~\cite{ref:boss} incorporating the detector calibration, event reconstruction and data storage.

\section{event selection}
Charged tracks in the polar angle range $|\rm{cos}\theta| < 0.93$ are reconstructed from hits in the MDC. 
Tracks must extrapolate to within 10 cm of the interaction point in the beam direction and 1 cm in the plane perpendicular to the beam. 
Each track is assumed to be a pion and no particle identification is applied. 
Candidate events are required to have two charged tracks and zero net charge.

Photon candidates are required to have an energy deposition above 25 MeV in the barrel region ($|\cos\theta|<0.80$) and 50 MeV in the end cap ($0.86<|\cos\theta|<0.92$). To exclude showers from charged tracks, the angle between the shower position and the charged tracks extrapolated to the EMC must be greater than $10^{\circ}$. A timing requirement in the EMC is used to suppress electronic noise and energy deposits unrelated to the event. At least six (seven) photons are required for the $\etap\to\gammapipi$ ($\etap\to\pipieta$) mode.

\begin{figure}[htbp]
\centering
\includegraphics[width=0.24\textwidth]{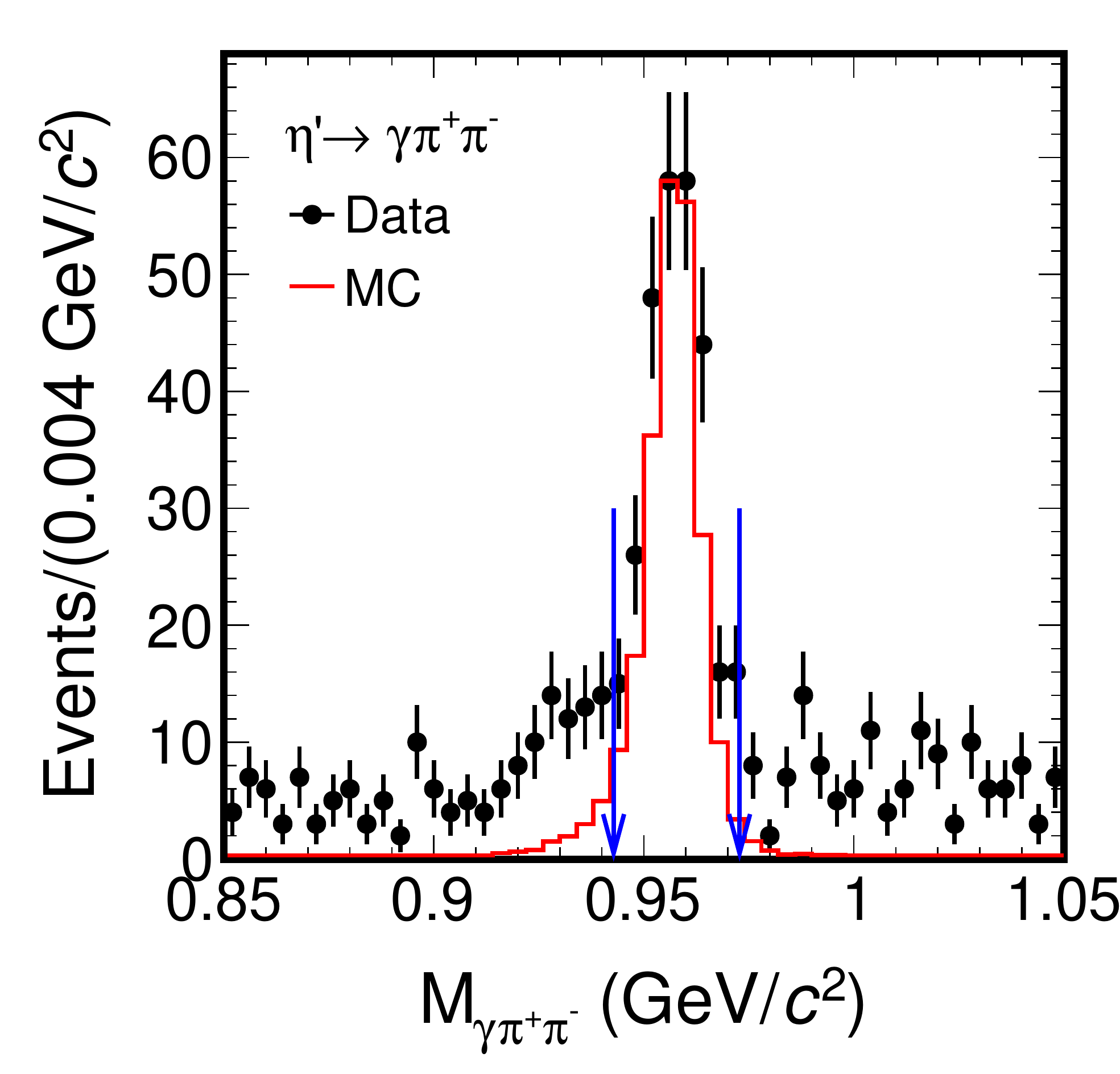}\put(-35,80){{(a)}}
\includegraphics[width=0.24\textwidth]{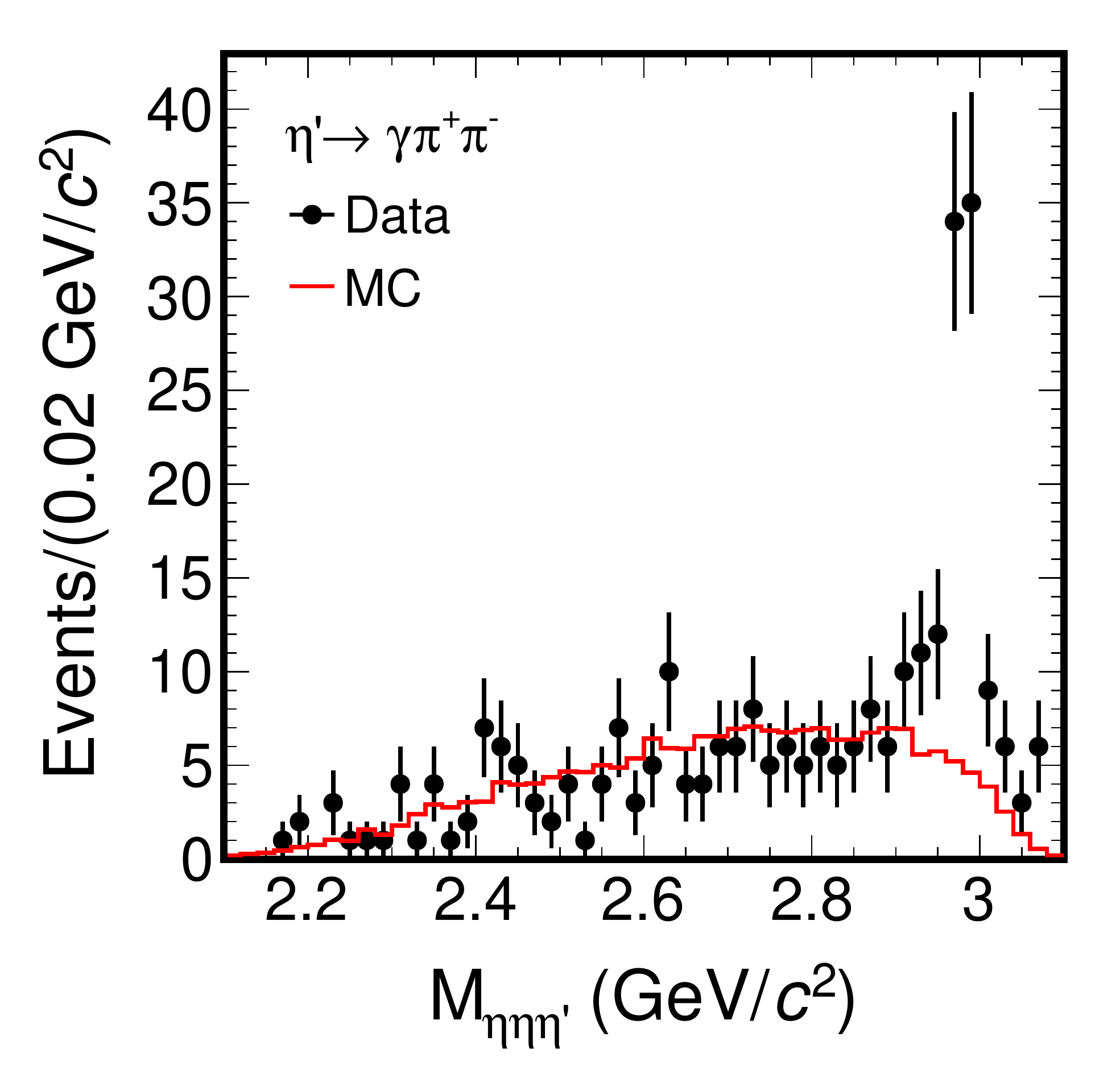}\put(-35,80){{(b)}}
\vskip -0cm
\includegraphics[width=0.24\textwidth]{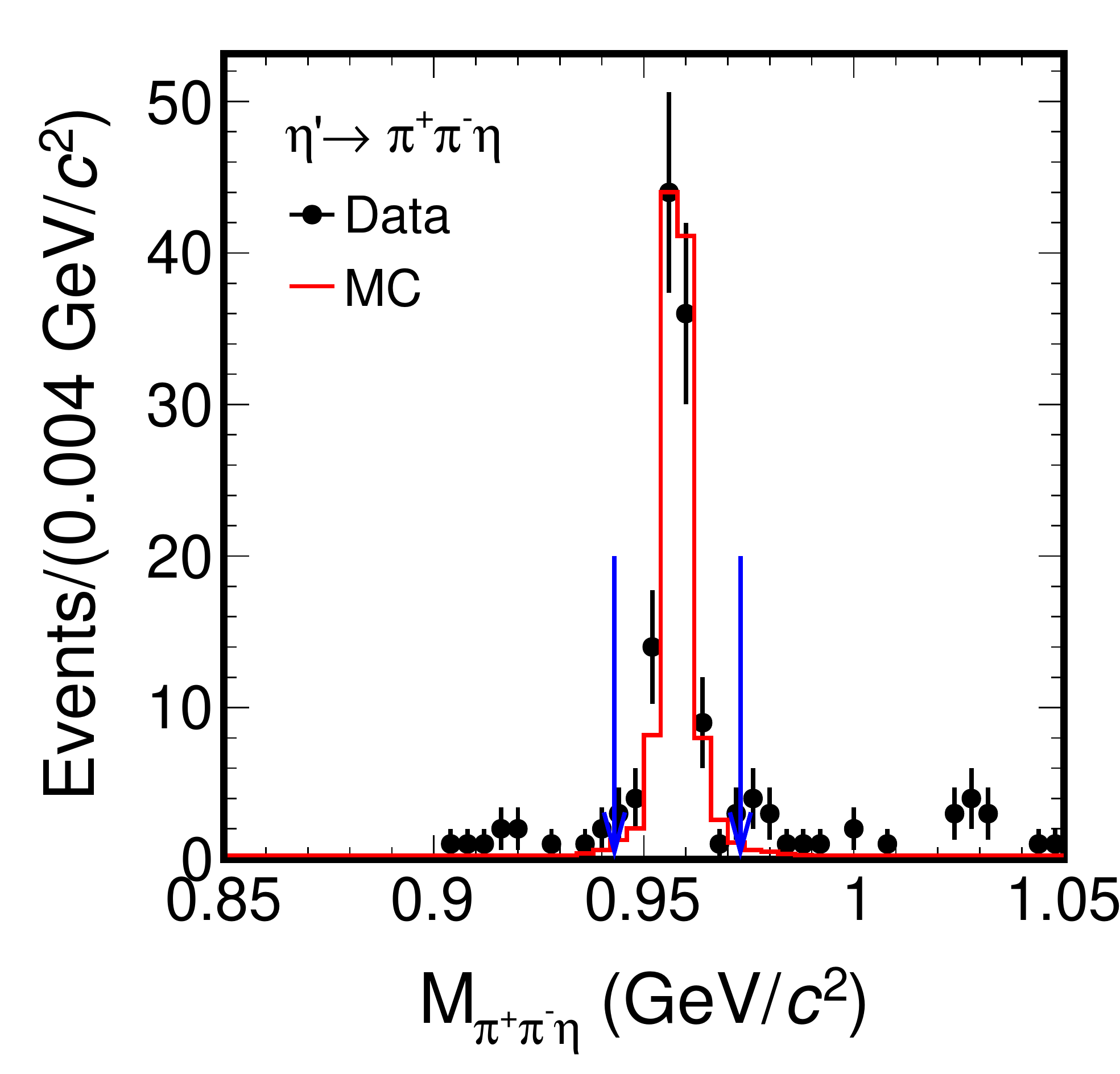}\put(-35,80){{(c)}}
\includegraphics[width=0.24\textwidth]{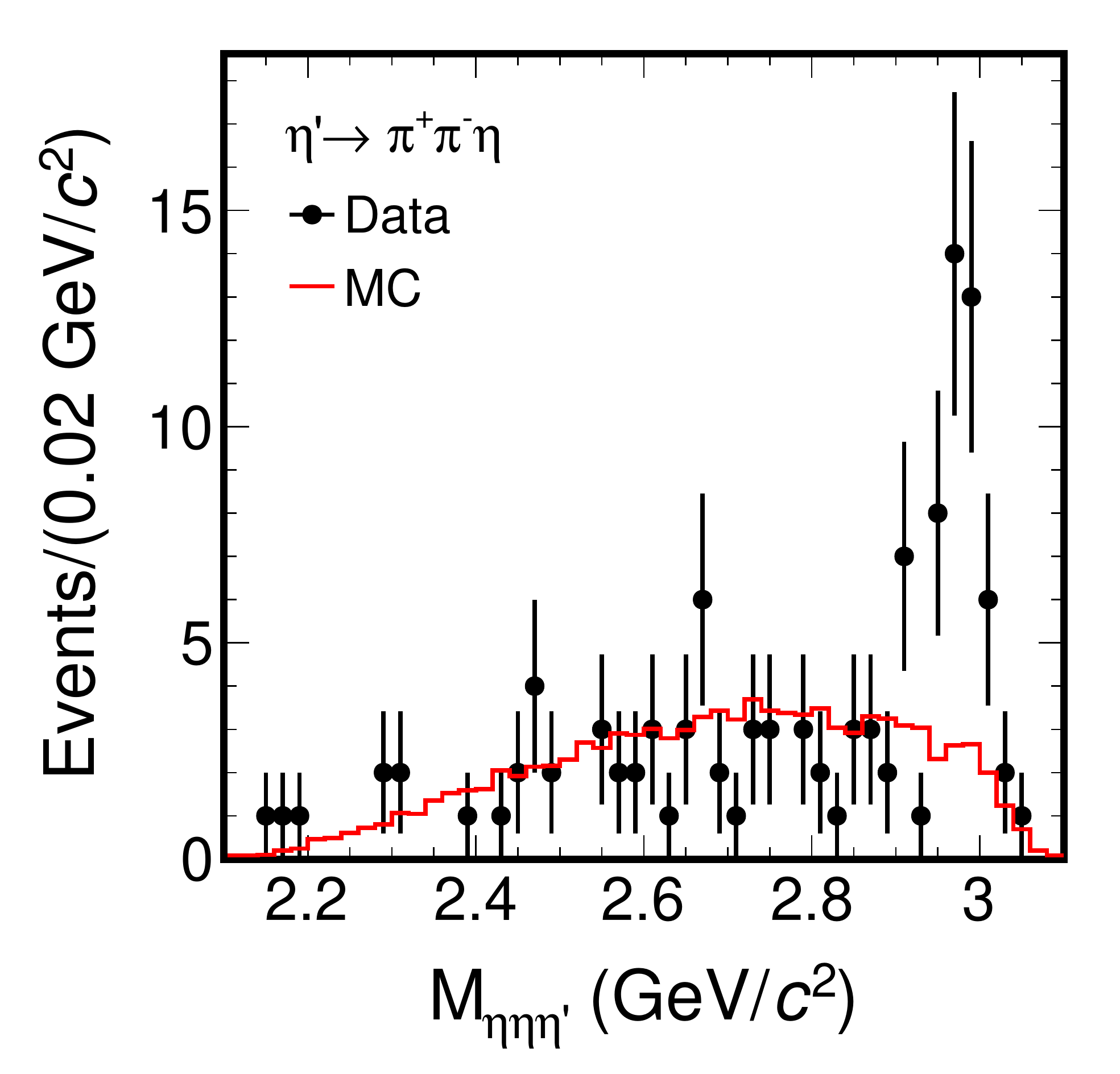}\put(-35,80){{(d)}}

\caption{\label{selectkketap}
     Invariant mass distributions for the selected candidates of $\jpsi\to\gamma\eta\eta\etap$.
     Plots (a) and (b) are the invariant mass distributions of $\gammapipi$
     and $\eta\eta\etap$ for $\etap\to\gamma\pp$, respectively;
     (c) and (d) are the invariant mass distributions of $\pipieta$
     and $\eta\eta\etap$ for $\etap\to\pipieta$, respectively.
      The dots with error bars are data and the histograms are for the signal MC samples (arbitrary normalization).
      }
\end{figure}

For the $\jpsi\to\gamma\eta\eta\etap$, $\etap\to\gamma\pp$ channel, a six-constraint (6C)  kinematic fit is performed to the hypothesis of $\jpsi\to\gamma\gamma\eta\eta\pip\pim$. 
This includes a 4C fit to the $\jp$ initial four-momentum and 1C fit of each pair of photons to have an invariant mass equal to that of an $\eta$. 
For events with more than six photon candidates, the combination with the minimum $\chi_{6C}^{2}$ is selected, and $\chi_{\rm 6C}^{2}<$ 30 is required.
Events with $|M_{\gamma\gamma} - m_{\pi^{0}}| <$ 0.02~GeV/$c^{2}$ are rejected to suppress background containing a $\pi^{0}$, where the $m_{\pi^{0}}$ is the nominal mass of the $\pi^{0}$~\cite{pdg}.
In order to reduce the background due to mis-reconstruction of the event, events with $|M_{\tilde{\gamma}\tilde{\gamma}} - m_{\eta}| <$ ~0.02~GeV/$c^{2}$ are rejected, where the $M_{\tilde{\gamma}\tilde{\gamma}}$ is the invariant mass of all photon pairs except the pairs from the constrained $\eta$ candidates and $m_{\eta}$ is the nominal mass of $\eta$~\cite{pdg}.
A clear $\etap$ signal is observed in the invariant mass distribution of $\gammapipi$ ($M_{\gamma\pi^{+}\pi^{-}}$), as shown in Fig.~\ref{selectkketap}(a).
The $\pp$ invariant mass is required to be near the $\rho$ mass region, $M_{\pi^{+}\pi^{-}} >$ 0.5~GeV/$c^{2}$.
Candidate $\etap$ is reconstructed from the $\gammapipi$ pair with $|M_{\gamma\pi^{+}\pi^{-}} - m_{\etap}| <$ 0.015~GeV/$c^{2}$, where the $m_{\etap}$ is the nominal mass of the $\etap$~\cite{pdg}.
If there is more than one combination, we select the one with $M_{\gamma\pi^{+}\pi^{-}}$ closest to $m_{\etap}$.

After applying the requirements above, we obtain the invariant mass distribution of $\eta\eta\etap$ ($M_{\eta\eta\etap}$), in which clear $\etac$ signal is observed, as shown in Fig.~\ref{selectkketap}(b).

For the $\jpsi\to\gamma\eta\eta\etap, \etap\to\pipieta$ channel, a seven-constraint (7C) kinematic fit is performed to the hypothesis of $\jpsi\to\gamma\eta\eta\eta\pip\pim$ in order to improve the $\etap$ mass resolution.
If there are more than seven photon candidates, the combination with the minimum $\chi_{7C}^{2}$ is retained, and $\chi_{\rm 7C}^{2} < $ 50 is required. 
To suppress background from $\pi^{0}\to\gamma\gamma$,
$|M_{\gamma\gamma} - m_{\pi^{0}}| >$ 0.02~GeV/$c^{2}$ is required for all photon pairs. 
In order to reduce the background due to wrong reconstruction of the event, events with $|M_{\gamma_{r}\gamma_{\eta}} - m_{\eta}| <$~0.02~GeV/$c^{2}$ are rejected, where the $M_{\gamma_{r}\gamma_{\eta}}$ is the invariant mass of the radiative photon ($\gamma_{r}$) directly from $\jpsi$ decays paired with any photon from an $\eta$ candidate decay ($\gamma_{\eta}$).
The $\etap$ candidates are formed from $\pipieta$ combination satisfying $|M_{\pi^{+}\pi^{-}\eta} - m_{\etap}| <$ 0.015 GeV/$c^{2}$ and the combination with $M_{\pi^{+}\pi^{-}\eta}$ closest to $m_{\etap}$ is selected, where $M_{\pi^{+}\pi^{-}\eta}$ is the invariant mass of $\pi^{+}\pi^{-}\eta$, as shown in Fig.~\ref{selectkketap}(c).
Finally, the invariant mass distribution of $\eta\eta\etap$ ($M_{\eta\eta\etap}$), with a clear signal of $\etac$, is shown in Fig.~\ref{selectkketap}(d).

\section{Signal extraction}
Potential backgrounds are studied using an inclusive MC sample of $1.2\times10^{9}$ $\jpsi$ decays.
No significant peaking background is observed in the invariant mass distribution of $\eta\eta\etap$.
Non-$\etap$ processes are studied using the $\etap$ mass sidebands, which are [0.890, 0.920] and [0.995, 1.025] GeV/$c^{2}$.
No clear peak is observed in $X(2370)$ and $\etac$ mass region from sideband study.

\begin{figure*}[htbp!]
\centering
	\includegraphics[width=0.46\textwidth]{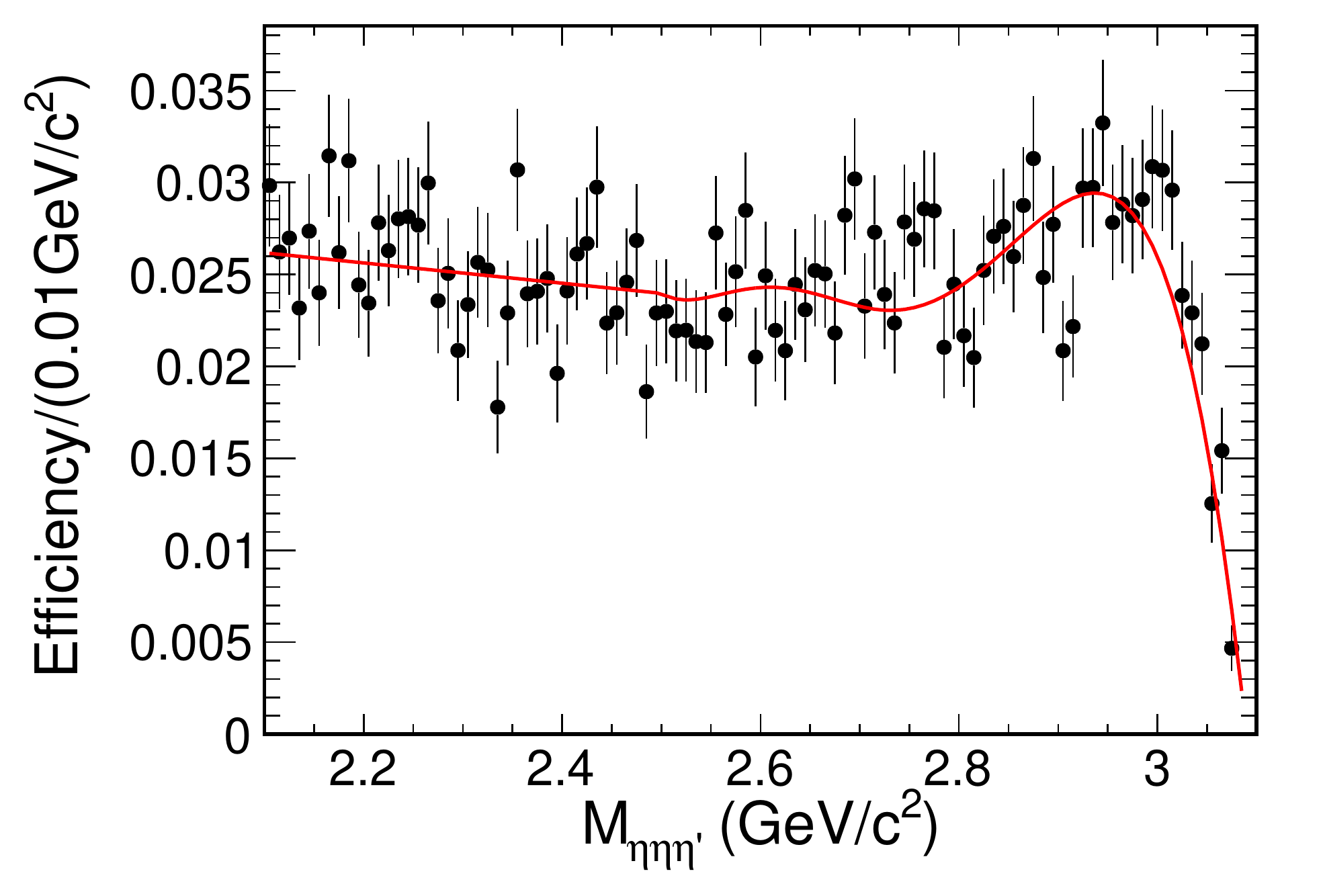}
	\put(-128,138){{(a)}} 
	\includegraphics[width=0.46\textwidth]{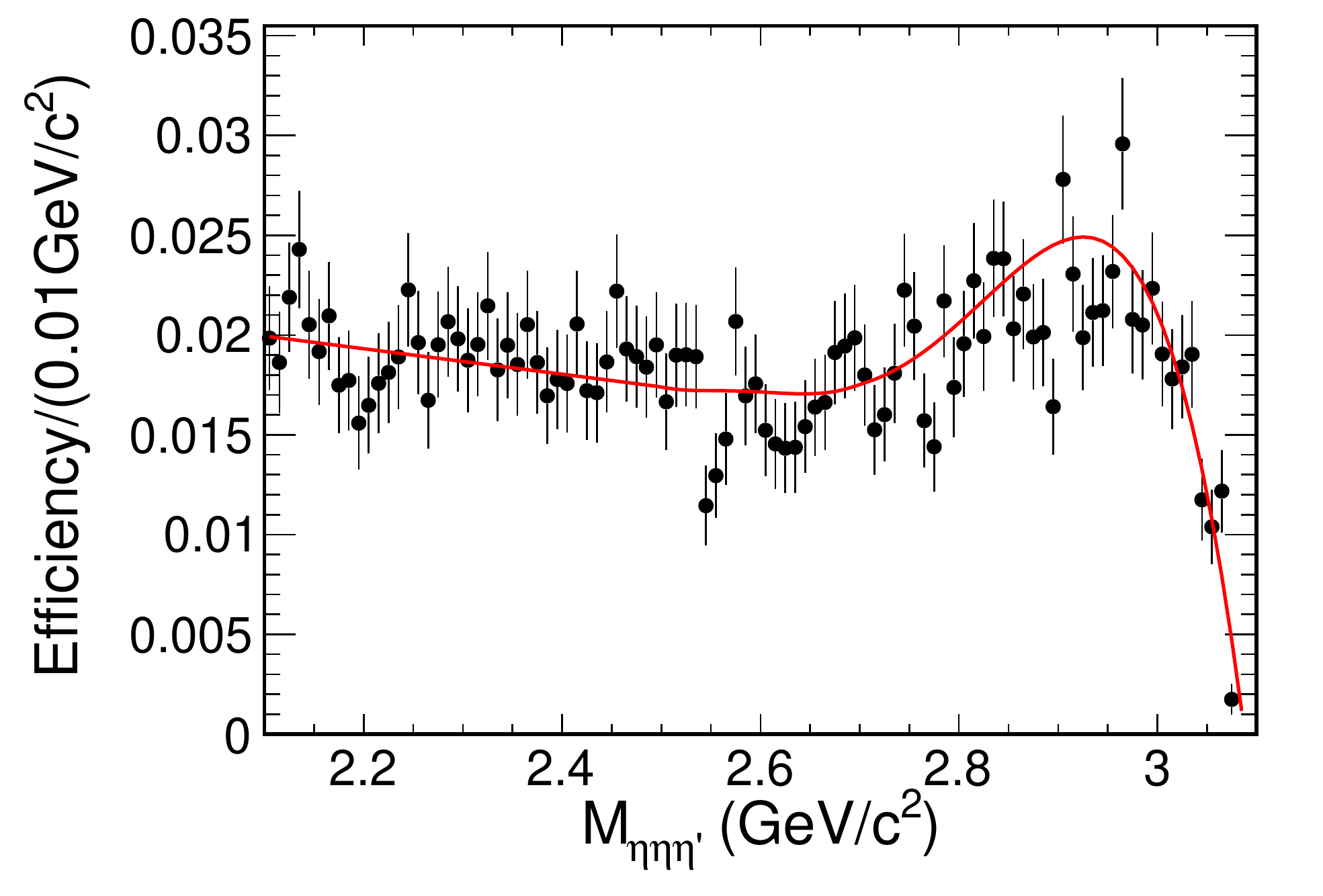}
	\put(-128,138){{(b)}} 
		\vskip -0cm
	\includegraphics[width = 0.45\textwidth]{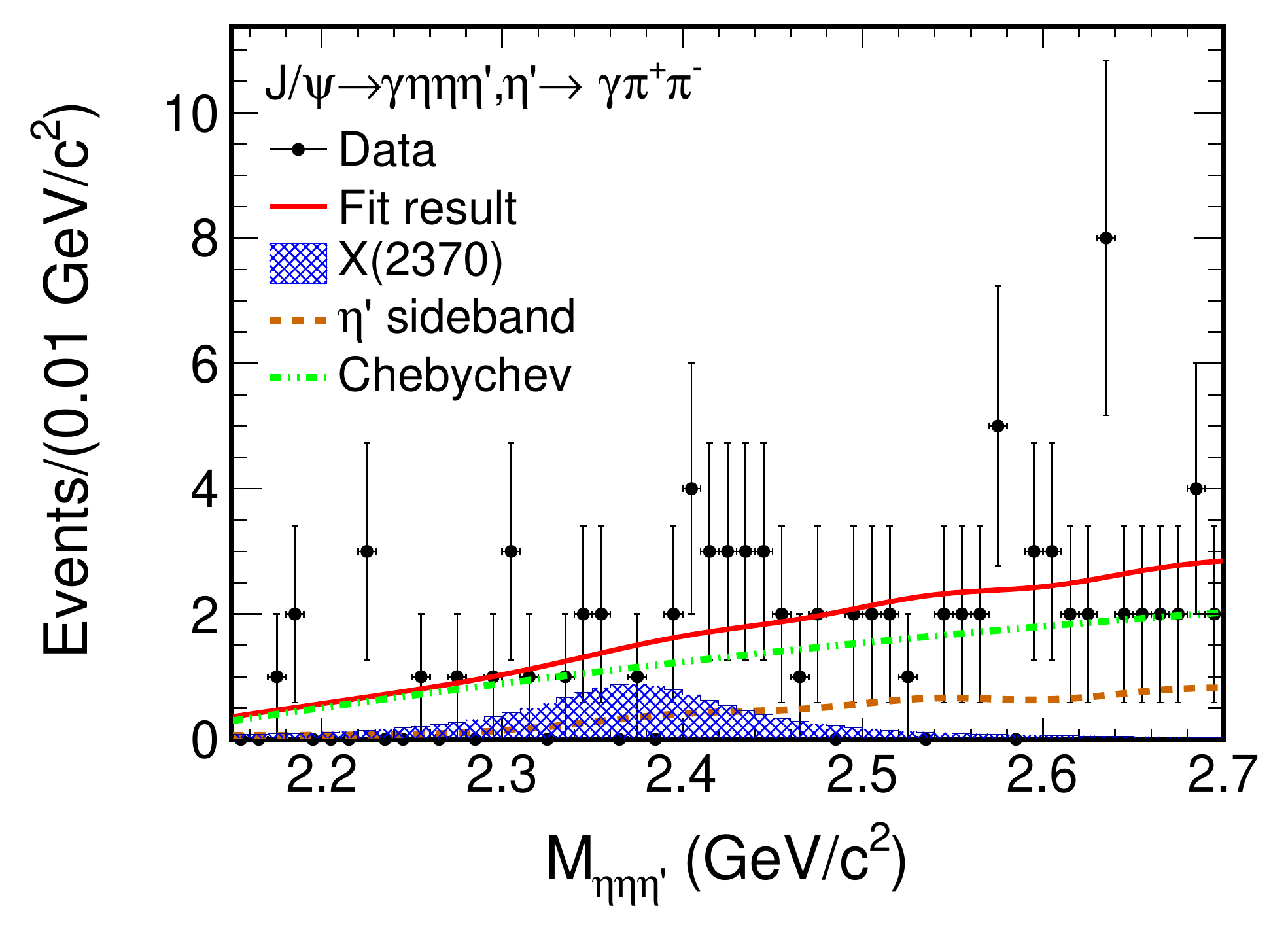}
	\put(-128,138){{(c)}} 
	\includegraphics[width = 0.45\textwidth]{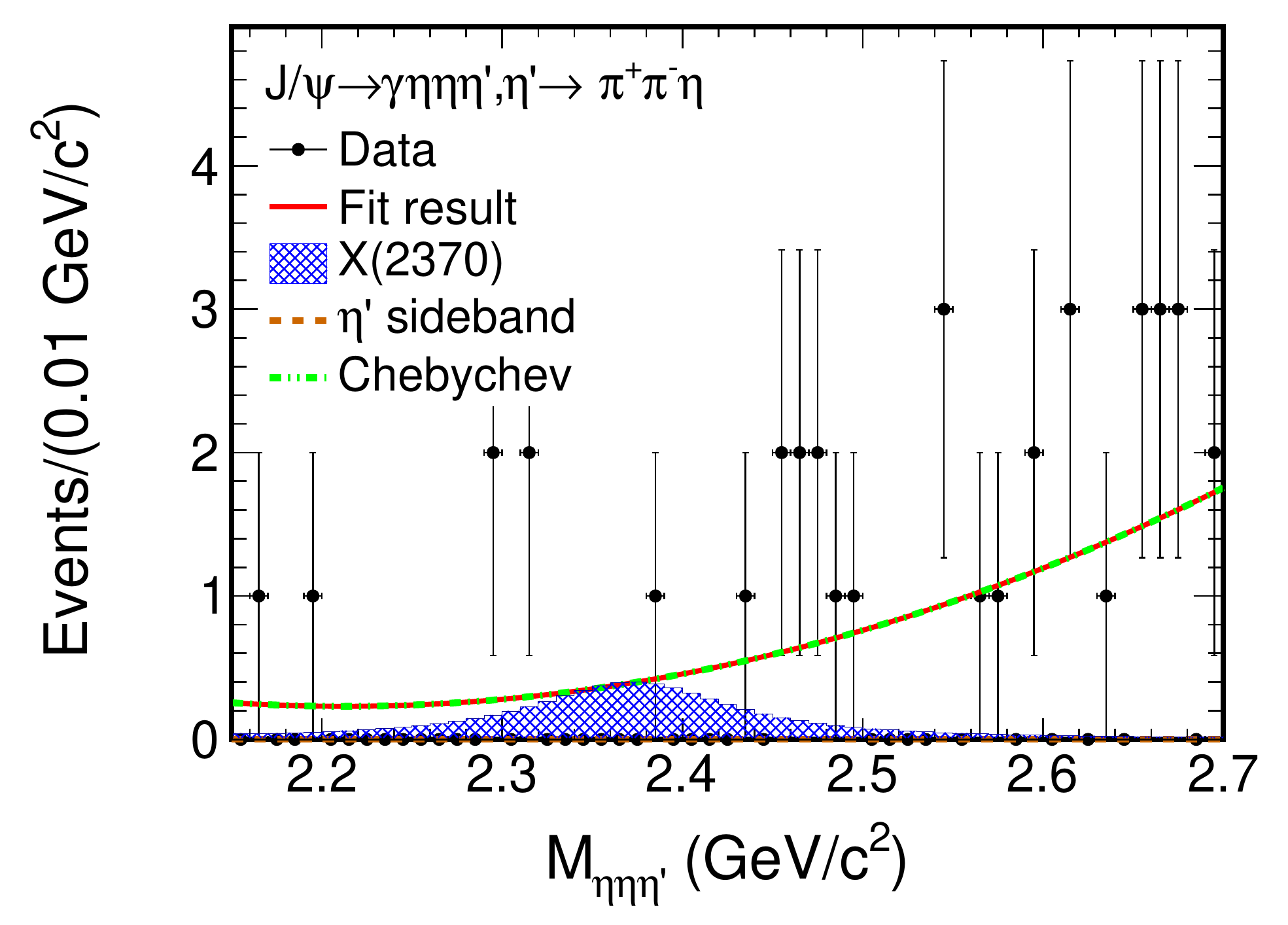}
	\put(-128,138){{(d)}}
		\vskip -0cm
	\includegraphics[width=0.45\textwidth]{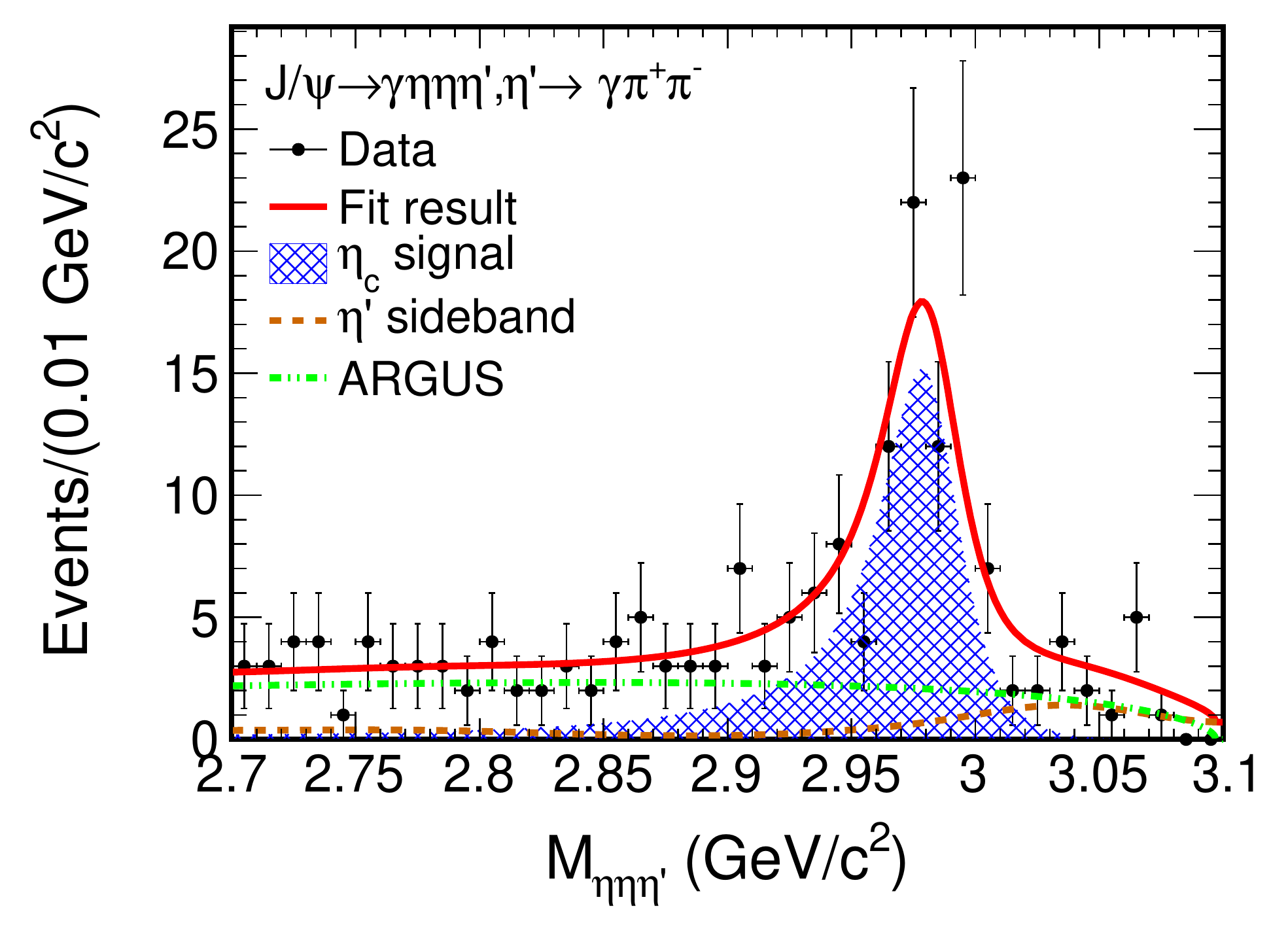}
	\put(-128,138){{(e)}}
	\includegraphics[width=0.45\textwidth]{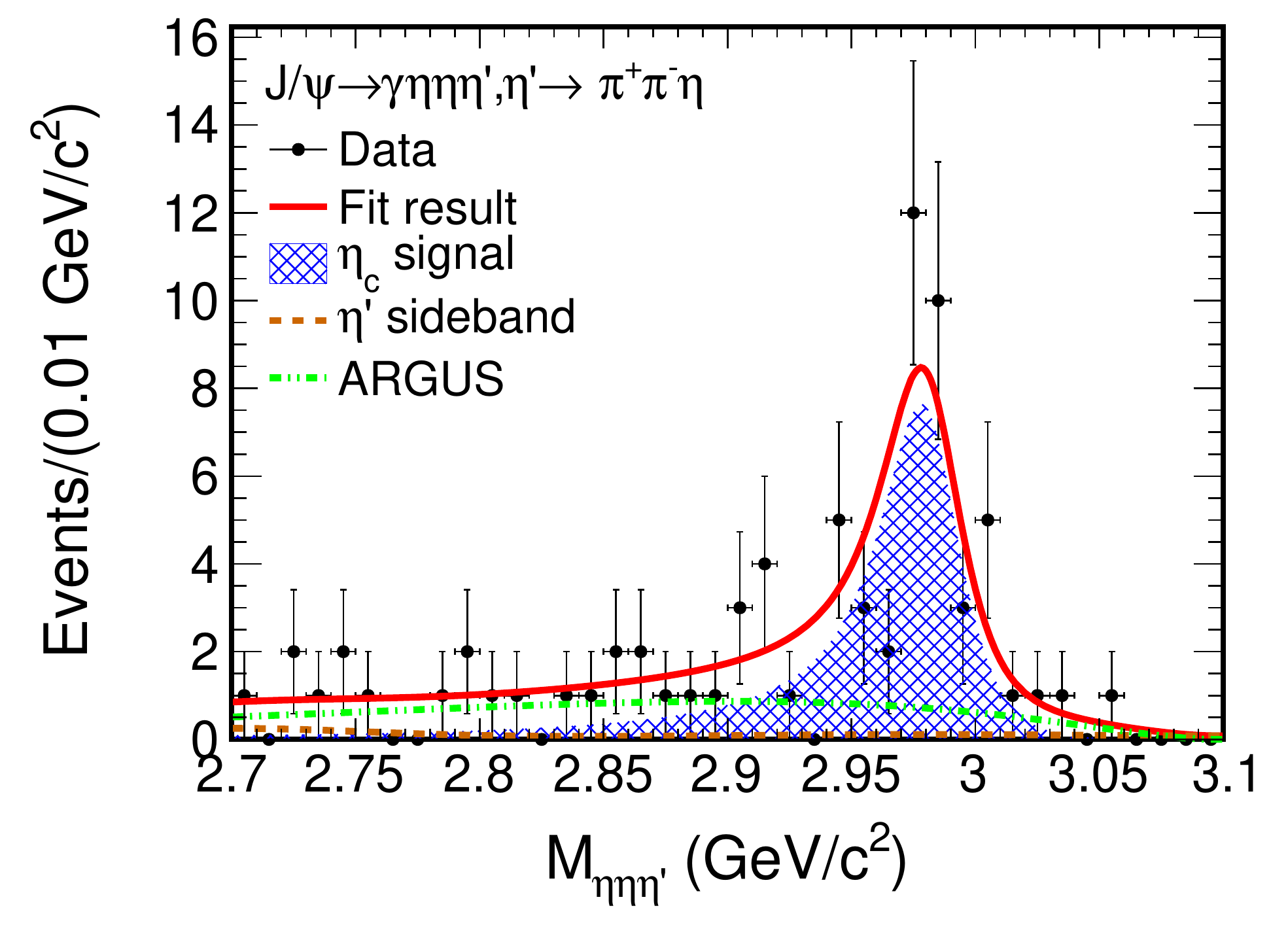}
	\put(-128,138){{(f)}}
\caption{\label{fig:eff_fit}
     Plots (a) and (b) are efficiency curves for the decays of $\etap\ra\gam\pp$ and $\etap\ra\pp\eta$ obtained from $\jpsi\to\gamma X \ra \gam \eta\eta\etap$ MC simulation, where X means $0^{-+}$ non-resonant state.
     Plots (c) and (d) are the simultaneous fit results for the $X(2370)$ in the invariant mass distribution of $\eta\eta\etap$ for the decays of $\etap\to\gammapipi$ and $\etap\to\pipieta$, respectively.
     Plots (e) and (f) are the fit results for $\etac$ in the invariant mass distribution of $\eta\eta \etap$ for the decays of $\etap\to\gammapipi$ and $\etap\to\pipieta$, respectively.
     The dots with error bars represent the data;
     the red solid curves show the fit results;
     the hatched areas represent the signal of the $X(2370)$ scaled to the upper limit or the signal of the $\etac$; 
     the brown dashed lines show the events from $\etap$ sideband;
     the green hyphenated lines represent the Chebychev polynomial function or the ARGUS function.
    }
\end{figure*}

Efficiency curves obtained from $0^{-+}$ PHSP MC simulation are shown in Figs.~\ref{fig:eff_fit}(a) and (b).
Using double Gaussian function to fit the invariant spectrum of $\eta\eta\etap$ from signal MC samples generated with a zero width resonance, the mass resolutions of the $X(2370)$ in these two $\etap$ decay modes are determined to be 8.2 MeV/$c^{2}~(\etap\ra\gam\pp)$ and 8.7 MeV/$c^{2}~(\etap\ra\pp\eta)$; while the mass resolutions of the $\etac$ are determined to be 5.4 MeV/$c^{2}~(\etap\ra\gam\pp)$ and 5.7 MeV/$c^{2}~(\etap\ra\pp\eta)$.

There is no obvious signal for the $X(2370)$ in $\eta\eta\etap$ invariant mass distributions in Figs.~\ref{selectkketap}(b) and (d).  
We perform a simultaneous unbinned maximum likelihood fit to the $\eta\eta\etap$ distributions in the range of [2.1, 2.7] GeV/$c^{2}$. 
The results are shown in Figs.~\ref{fig:eff_fit}(c) and (d), where the signal size represents the upper limit fit result of the $X(2370)$ rather than the negligible central value from the actual fit. 
The $X(2370)$ signal peak is represented by an efficiency-weighted non-relativistic Breit-Wigner (BW) function convolved with a double Gaussian function to account for the mass resolution.
Due to low statistics, the mass and width of the BW function are fixed to previously published BESIII results~\cite{PRL1} while the parameters of the double Gaussian function are fixed to the results obtained from the fit of signal MC samples generated with zero width of the $X(2370)$.
Interference between the $X(2370)$ and other components is ignored.
The non-$\etap$ background events are described using $\etap$ mass sidebands and the yields are fixed in the fit; the remaining background is described by a second order Chebychev polynomial function with free parameters.
In the simultaneous fit, the signal ratio for the two $\etap$ decay modes are fixed with a factor calculated by their branching fractions and efficiencies.
Since no evident $X(2370)$ signal is seen in $M_{\eta\eta\etap}$, a Bayesian method is used to obtain the upper limit of the signal yield at the 90\% confidence level (C.L.).
To determine the upper limit of the signal yield, the distribution of normalized likelihood values for a series of expected signal event yields is taken as the probability density function (PDF).
The 90\% C.L. yield, $N^{UL}$, is set such that 90\% of the PDF area above zero yield is contained between 0 and $N^{UL}$.  We repeat this procedure with different $X(2370)$ signal shape parameters, fit ranges, $\etap$ sideband regions and background shapes, and the maximum upper limit among these cases is selected. 
The obtained upper limits of the signal yields are listed in Table~\ref{sum_yields}. 
The MC detection efficiencies of $\jpsi\ra\gam X(2370)\ra\gam\doube\etap$ for the two $\etap$ decay modes are determined to be 2.95\% ($\etap\ra\gam\pp$) and 2.32\% ($\etap\ra\pp\eta$).
The upper limit of the product branching fraction is $\cal{B}$ $(\jpsi\ra\gam X(2370)\cdot{\cal B}(X(2370)\ra\doube\etap) <$ $8.70 \times 10^{-6}$.

\begin{table}[htbp!]
	\begin{center}
		{\caption{ Fit results of the signal yields for $\jpsi \ra \gam X(2370) \ra \gam \eta\eta\etap$ and $\jpsi \ra \gam \etac \ra \gam \eta\eta\etap$. The uncertainties are statistical only.}
			\label{sum_yields}}		
		\begin{tabular}{lcc}  
		\hline\hline			
		    Decay channel                    &$\etap\to\gammapipi$ &$\etap\to\pipieta$ \\ \hline            
            $\jpsi \ra \gam X(2370) \ra \gam \eta\eta\etap$ & $< 15.2$   & $< 6.9$ \\
            $\jpsi \ra \gam \etac \ra \gam \eta\eta\etap$  & 93.3$\pm$11.9  & 43.2$\pm$5.5 \\  
	 \hline \hline
		\end{tabular}
	\end{center}
\end{table}

A clear signal for the $\etac$ is observed in $\eta\eta\etap$ invariant mass distributions. 
We perform a simultaneous unbinned maximum likelihood fit to the $\eta\eta\etap$ distributions in the range of [2.70, 3.10] GeV/$c^{2}$, as shown in Figs.~\ref{fig:eff_fit}(e) and (f).
The $\etac$ signal is described with an efficiency-weighted $E_{\gamma}^{3}\times f_{\rm damp}(E_{\gamma})\times BW(m)$ function convolved with a double Gaussian function, where $m$ is the $\eta\eta\etap$ invariant mass and $E_{\gamma}=\frac{m_{J/\psi}^{2}-m^{2}}{2m_{J/\psi}}$ is the energy of the transition photon in the rest frame of $J/\psi$.  
We also insert the function $f_{\rm damp}(E_{\gamma})=\frac{E_{0}^{2}}{E_{0}E_{\gamma}+(E_{0}-E_{\gamma})^{2}}$ 
to damp the divergent tail at low mass arising from the $E_{\gamma}^{3}$ behavior, where $E_{0}=\frac{m_{J/\psi}^{2}-m_{\eta_{c}}^{2}}{2m_{J/\psi}}$ is the nominal energy of the transition photon~\cite{KEDRetac}.
The mass and width of the $\etac$ are fixed to PDG values~\cite{pdg}. 
Interference between $\etac$ and other components is ignored.
Backgrounds are modeled with similar components as for the fit of the $X(2370)$ discussed above, while the Chebychev polynomial is replaced with an ARGUS function~\cite{ARGUS}.
The obtained signal yields, which have correlated uncertainties due to the constrianed fit, for $\jp\to\gamma \etac\to\gamma\eta\eta\etap$ are listed in Table~\ref{sum_yields}.
The detection efficiencies of $\jpsi\ra\gam \etac\ra\gam\doube\etap$ for two $\etap$ decay modes are determined to be 
2.94\% ($\etap\ra\gam\pp$) and 2.35\% ($\etap\ra\pp\eta$).
We observe some disagreements in the data vs.~MC simulated $\eta\eta$, $\eta\etap$, and $\eta\eta\etap$ invariant mass spectra.  
We employ a machine learning (ML) method~\cite{ML} 
to re-weight the signal MC events based on the meson candidate's four-momenta.  
This reduces the inconsistency between data and signal MC, providing an accurate efficiency.  
The product branching fraction of $\jp\to\gamma \etac\to\gamma \eta\eta \etap$ is then determined to be $(4.86\pm0.62(\rm stat.))\times10^{-5}$. 
The statistical significance of $\etac$ is determined to be 8.1$\sigma$.

\section{SYSTEMATIC UNCERTAINTIES} \label{sec::sys}

Several sources of systematic uncertainties are considered, including the data-MC efficiency differences in the MDC tracking and the photon detection efficiency,  the kinematic fit, and the mass window requirements for the $\pi^{0}$, $\eta$, $\rho$ and $\etap$.
Uncertainties associated with the fit ranges, the background shapes, the sideband regions, quantum number of $X(2370)$, the signal shape parameters of $\etac$, damping factor, efficiency calculation, intermediate resonance decay branching fractions and the total number of $\jpsi$ events are considered.

\begin{table}[htp]
	\begin{center}
		{\caption {Systematic uncertainties for determination of the upper limit of branching fraction of $\jp\to\gamma X(2370)\to\gamma \eta\eta \etap$ (in \%).
The items with * are common uncertainties of both  $\etap$ decay modes. }
			\label{sumsysBR2370}}
		
		\begin{tabular}{lcc}  
		\hline\hline
			
		    Source                                    &$\etap\to\gammapipi$ &$\etap\to\pipieta$ \\ \hline

            MDC tracking*                             &2.0   &2.0\\
            Photon detection*                         &6.0   &7.0\\
			Kinematic fit                             &1.0   &1.0\\
			$\rho$ mass window                        &2.4   &--\\
			$\etap$ mass window                       &1.2   &0.6\\	
			$\pi^{0}$ veto                            & 18.6   &5.3 \\
			$\eta$ veto                              & 15.5   &0.6 \\	
            Quantum numbers of $X(2370)$                     &13.4  &10.5 \\
            $B(\etap\to \gamma\pippim)$               &1.7   &--\\
			$B(\etap\to\pippim\eta)$                  &--    &1.6 \\
			$B(\eta\to\gamma\gamma)$*                 & 1.0   &1.5\\	
            Number of $\jpsi$ events*                 &0.5   &0.5\\
			Total                                     &28.6  &14.1\\ \hline \hline
		\end{tabular}
	\end{center}
\end{table}

\begin{table}[htp]
	\begin{center}
		{\caption {Systematic uncertainties for the determination of the branching fraction of $\jp\to\gamma \etac\to\gamma \eta\eta \etap$(in \%).
The items with * are common uncertainties of both $\etap$ decay modes.}
			\label{sumsysBRetac}}

		\begin{tabular}{lcc}  \hline\hline
			
		    Source                                    &$\etap\to\gammapipi$ &$\etap\to\pipieta$ \\ \hline

            MDC tracking*                             &2.0   &2.0\\
            Photon detection*                         &6.0   &7.0\\
			Kinematic fit                             &1.0  &1.0\\
			$\rho$ mass window                        &1.6   &--\\
			$\etap$ mass window                       &0.1   &0.1\\	
			$\pi^{0}$ veto                            &0.6   &1.5 \\
			$\eta$ veto                               &1.2   &0.0\\			
            Efficiency calculation with ML            &1.4   &1.4\\
			Fit range                                 &1.9   &1.9 \\
            Sideband region                           &0.0   &2.2 \\
			Background shape                       	&11.7   &5.9 \\	
            $B(\etap\to \gamma\pippim)$               &1.7  &--\\
			$B(\etap\to\pippim\eta)$                  &--   &1.6 \\
			$B(\eta\to\gamma\gamma)$*                 &1.0   &1.5\\	
            Number of $\jpsi$ events*                 &0.5  &0.5\\
            Parameters of $\etac$                     &2.8  &2.8\\
            Damping factor                            &1.7  &1.7\\
			Total                                     &14.2  & 10.8 \\ \hline \hline
		\end{tabular}
		
	\end{center}
\end{table}

\subsection{Efficiency estimation}
The MDC tracking efficiencies of charged pions are investigated using a clean control sample of $\jp\to p\bar{p}\pippim$~\cite{MDCpi}. 
The difference in tracking efficiencies between data and MC simulation is 1.0\% for each charged pion.
The photon detection efficiency is studied with a clean sample of
$J/\psi\to\rho^{0}\pi^{0}$~\cite{Photon}. The result shows that
the data-MC efficiency difference is 1.0\% per photon.

The systematic uncertainties associated with the kinematic fit are studied with the track helix parameter correction method, as described in Ref.~\cite{4cError}. 
The differences with respect to those without corrections are taken as the systematic uncertainties.

Due to the difference in the mass resolution between data and MC simulation, the uncertainties related to the $M_{\pp}$ and $\etap$ mass window requirements are investigated by smearing the MC simulation to improve the consistency between data and MC simulation. 
The differences of the detection efficiency before and after smearing are assigned as  systematic uncertainties for the $M_{\pp}$ and $\etap$ mass window requirements.
The uncertainties from the $\pi^{0}$ and $\eta$ mass window requirements are estimated by varying those mass windows.
The changes in the resultant branching fractions are assigned as the systematic uncertainties from these items.

To study uncertainties related to the efficiency calculation with the ML method, we generate a generic MC sample with $\jp\to\gamma\eta_{c},\eta_{c}\to f_{2}(1810)\etap(f_{2}(1810)\to\eta\eta)$ process to represent the signal and $\jp\to\gamma\eta\eta\etap$ as the non-$\eta_{c}$ background. 
The numbers of signal and background events are fixed to fitting results. 
The efficiency difference between the generic MC sample and the ML method are taken as systematic uncertainty from this item.
Furthermore, we consider the effects arising from different quantum numbers of the $X(2370)$. We generate $\jp\rightarrow\gamma X(2370)$ decays under the assumption of a $\mathrm{sin}^2\theta_{\gamma}$ angular distribution. The resulting difference of efficiency with respect to the nominal value is taken as systematic uncertainty.

\subsection{Fit to the signal}
Systematic uncertainties related to the $X(2370)$ signal treatment are 
already accounted for in the upper limit yield, as discussed previously; 
here, we discuss the treatment of the $\etac$ signal.  
To study the uncertainties from the fit range, the fits are repeated with different fit ranges, and the largest difference among these signal yields is taken as systematic uncertainty.
The uncertainties from the $\etap$ sideband region are estimated by using alternative sideband regions.
The maximum difference among signal yields with respect to the nominal value is taken as the uncertainty.
To estimate the uncertainty associated with the background shape, alternative fit with a truncated second order polynomial for the background is performed. 
The maximum difference in signal yields with respect to the nominal value is taken as systematic uncertainty.
To study the uncertainty associated with the parameters of $\etac$, we change these values by $\pm 1\sigma$ and repeat the fit. 
The largest difference from our nominal result among these alternative fits is taken as the uncertainty. 
The uncertainty due to damping factor is estimated by using an alternative form of the damping factor, which was used by the CLEO collaboration~\cite{CLEOdamp}, $f_{\rm damp}(E_{\gamma})=\mathrm{exp}\left(-\frac{E_{\gamma}^{2}}{8\beta^{2}}\right)$, 
where $E_{\gamma}$ is the energy of the transition photon and $\beta$ = 0.065 GeV. 
The difference between the results with different damping factor forms is taken as the systematic uncertainty.

\subsection{Other Uncertainties}
The uncertainties on the intermediate decay branching fractions of $\etap\to\gamma\pi^{+}\pi^{-}$, $\etap\to\pi^{+}\pi^{-}\eta$ and $\eta\to\gamma\gamma$ are taken from the world average values~\cite{pdg}, which are  $1.7\%$, $1.6\%$ and $0.5\%$, respectively.
The systematic uncertainty due to the number of $\jp$ events is determined as 0.5$\%$ according to Ref.~\cite{jpsinumber}.

A summary of all the uncertainties are shown in Tables~\ref{sumsysBR2370} and \ref{sumsysBRetac}.
The total systematic uncertainties are obtained by adding all individual uncertainties in quadrature, assuming all sources to be independent.

In this paper, $\jpsi\to\gamma \eta\eta\etap$ is studied with two $\etap$ decay modes.
The measurements from the two $\etap$ decay modes are, therefore, combined by considering the difference of uncertainties for these two measurements.
The combination of common and independent systematic uncertainties for the two $\etap$ decay modes are calculated with weighted least squares method~\cite{combinepaper}. 
The total systematic uncertainties are 12.8\% and 9.2\% for $\cal{B}$$(\jp\to\gamma X(2370))\cdot{\cal B}(X(2370)\to\eta\eta \etap)$ and $\cal{B}$$(\jp\to\gamma \etac)\cdot{\cal B}(\etac\to\eta\eta \etap)$, respectively.

\section{RESULTS AND SUMMARY}

Using a sample of $1.31\times10^{9} ~J/\psi$ events collected with the BESIII detector, the decays of $J/\psi\to\gamma \eta\eta\etap$ are investigated using the two $\etap$ decay modes, $\etap\to\gamma\pi^{+}\pi^{-}$ and $\etap\to\pi^{+}\pi^{-}\eta$, $\eta\to\gamma\gamma$.

No evident signal for the $X(2370)$ is observed in the $\eta\eta\etap$ invariant mass distribution. 
To obtain the signal upper limit, we use the Bayesian method and perform unbinned maximum likelihood fits to the invariant mass spectrum of $\eta\eta\etap$ with a series of expected signal yields. 
The distribution of normalized likelihood values is taken as the PDF for the expected signal yields.
The final upper limit of the product branching fraction of $\jp\to\gamma X(2370)\to\eta\eta\etap$ incorporates the 12.8\% relative systematic uncertainty by convolving the likelihood distribution with a Gaussian function:
\begin{align}
L(N')=\int_{0}^{\infty}L(N)\frac{1}{\sqrt{2\pi}\sigma_{sys}}exp\left[{\frac{-(N'-N)^2}{2\sigma_{sys}^2}}\right]dN,
\end{align}
where $L(N)$ is the likelihood distribution, $\sigma_{sys}= 0.128 N$, and $N$ is the input signal yield. 
The resulting upper limit of $\cal{B}$$(\jp\to\gamma X(2370)\to\gamma \eta\eta \etap)$ is estimated to be $9.2\times10^{-6}$, which is not in contradiction with the value predicted in Ref.~\cite{PRD1} where $X(2370)$ is assumed as a pseudoscalar glueball.
To understand the nature of $X(2370)$, it is mandatory to measure its spin and parity and to search for it in more decay modes with higher statistics.

A clear $\etac$ signal is observed for the first time in the $\eta\eta\etap$ invariant mass spectrum, the product branching fraction of $\cal{B}$$(\jp\to\gamma \etac)\cdot{\cal B}(\etac\to\eta\eta \etap)$ is determined to be $(4.86\pm0.62~(\rm stat.)\pm0.45~(\rm sys.))\times10^{-5}$, which is compatible with the theoretical prediction of partial decay width of $\etac\ra\eta\eta\etap$ in Ref.~\cite{etac_predict}.

\begin{acknowledgments}

The BESIII collaboration thanks the staff of BEPCII and the IHEP computing center for their strong support. This work is supported in part by National Key Basic Research Program of China under Contract No. 2015CB856700; National Natural Science Foundation of China (NSFC) under Contracts Nos. 11625523, 11635010, 11675183, 11735014, 11822506, 11835012, 11922511, 11935015, 11935016, 11935018, 11961141012, 12061131003; 
the Chinese Academy of Sciences (CAS) Large-Scale Scientific Facility Program; Joint Large-Scale Scientific Facility Funds of the NSFC and CAS under Contracts Nos. U1732103, U1732263, U1832207; CAS Key Research Program of Frontier Sciences under Contracts Nos. QYZDJ-SSW-SLH003, QYZDJ-SSW-SLH040; 100 Talents Program of CAS; INPAC and Shanghai Key Laboratory for Particle Physics and Cosmology; ERC under Contract No. 758462; German Research Foundation DFG under Contracts Nos. 443159800, Collaborative Research Center CRC 1044, FOR 2359, FOR 2359, GRK 214; Istituto Nazionale di Fisica Nucleare, Italy; Ministry of Development of Turkey under Contract No. DPT2006K-120470; National Science and Technology fund; Olle Engkvist Foundation under Contract No. 200-0605; STFC (United Kingdom); The Knut and Alice Wallenberg Foundation (Sweden) under Contract No. 2016.0157; The Royal Society, UK under Contracts Nos. DH140054, DH160214; The Swedish Research Council; U. S. Department of Energy under Contracts Nos. DE-FG02-05ER41374, DE-SC-0012069.
\end{acknowledgments}

\end{document}